\documentclass[useAMS,usenatbib,usegraphicx]{mn2e}

\def\spose#1{\hbox to 0pt{#1\hss}}
\def\simlt{\mathrel{\spose{\lower 3pt\hbox{$\mathchar"218$}}
     \raise 2.0pt\hbox{$\mathchar"13C$}}}
\def\simgt{\mathrel{\spose{\lower 3pt\hbox{$\mathchar"218$}}
     \raise 2.0pt\hbox{$\mathchar"13E$}}}

\title[]{
The Stellar Halo Metallicity -- Luminosity Relationship\\ for Spiral Galaxies
}
\author[A. Renda, B.K. Gibson, M. Mouhcine, et~al.]{Agostino 
Renda$^{1}$\thanks{E--mail: arenda,bgibson@astro.swin.edu.au},
Brad K.  Gibson$^{1,6}$\footnotemark[1], 
Mustapha Mouhcine$^{2}$, Rodrigo A. Ibata$^{3}$,
\newauthor 
Daisuke Kawata$^{1,7}$\thanks{E--mail: dkawata@ociw.edu}, 
Chris Flynn$^{1,4}$, Chris B. Brook$^{1,5}$\vspace{0.2cm}\\
$^{1}$Centre for Astrophysics \& Supercomputing, Swinburne University,
Hawthorn, Victoria 3122, Australia\\
$^{2}$School of Physics and Astronomy, University of Nottingham,
University Park, Nottingham NG7 2RD, UK\\
$^{3}$Observatoire Astronomique de Strasbourg, 11, rue de l'Universit\' e,
67000 Strasbourg, France\\
$^{4}$Tuorlan Observatorio, V\" ais\" al\" antie 20,
21500 Piikki\" o, Finland\\
$^{5}$D\' epartement de Physique, de G\' enie Physique et d'Optique,
Universit\' e Laval, Qu\' ebec G1K 7P4, Canada\\
$^{6}$ School of Mathematical Sciences, Monash University, Clayton,
Victoria 3800, Australia\\
$^{7}$ The Observatories of the Carnegie Institution of Washington,
 813 Santa Barbara Street, Pasadena, CA 91101, USA
}

\begin{document}

\date{Accepted. Received; in original form}

\pagerange{\pageref{firstpage}--\pageref{lastpage}} \pubyear{0000}

\maketitle

\label{firstpage}

\begin{abstract}
The stellar halos of spiral galaxies bear important chemo--dynamical
signatures of galaxy formation. We present here the analysis of 89
semi--cosmological spiral galaxy simulations, spanning $\sim$~4~magnitudes
in total galactic luminosity. These simulations sample a wide variety of
merging histories and show significant dispersion in halo metallicity at a
given total luminosity - more than a factor of ten in metallicity. 
Our preliminary analysis suggests that galaxies
with a more extended merging history possess halos which have younger and
more metal rich stellar populations than the stellar halos associated with
galaxies with a more abbreviated assembly. A correlation between halo
metallicity and its surface brightness has also been found, reflecting the
correlation between halo metallicity and its stellar mass.  Our
simulations are compared with recent Hubble Space Telescope observations
of resolved stellar halos in nearby spirals.
\end{abstract}


\section{Introduction} 

Understanding the formation history of stellar halos is one of the
classical pursuits of galactic astronomy.  The problem is generally framed
within the context of two competing scenarios: one of ``rapid
collapse'' (Eggen, Lynden--Bell \& Sandage 1962), in which the stellar
halo is formed by the rapid collapse of a proto--galaxy within a dynamical
timescale ($\sim$~10$^{8}$~yr), and one of ``galactic assembly'' 
\citep{SZ}, whereby the stellar halo is assembled on a longer timescale
($\sim$~10$^{ 9}$~yr) by the accretion of ``building--blocks'', each with
separate enrichment histories.  Both scenarios have their strengths and
weaknesses, and it would appear that a hybrid model is the most plausible
option consistent with extant data (e.g.: \citealt{CB}; \citealt{FreemanBH}). 

An intriguing piece of the halo formation ``puzzle'' is provided by 
comparing the stellar halo of our own Milky Way with that of its
neighbour, M31. First, despite their comparable total galactic luminosities,
the stellar halo of M31 is {\it apparently} much more metal--rich than that of
the Milky Way (e.g.: \citealt{RN}; \citealt{MK}; Durrell, Harris \& Pritchet 2001; 
\citealt{Ferguson}; \citealt{Brown}; \citealt{Bellazzini}; \citealt{Durrell04}; \citealt{Ferguson05}; \citealt{Irwin}). In fact, the halo of M31 bears a closer resemblance to that of NGC~5128
(e.g., \citealt{HH}), despite their differing morphological classifications.

The stellar halo -- galaxy formation symbiosis has been further brought to
light by the recent work of Mouhcine et al. (see also Tikhonov, Galazutdinova \&
Drozdovsky 2005). The deep Hubble Space Telescope (HST hereafter) imaging of nearby spiral galaxy stellar
halos in \cite{MM} suggests a significant correlation between stellar 
halo metallicity and total galactic luminosity.  On the surface, this
correlation appears to leave our own Milky Way's halo as an
outlier with respect to other spirals, with a stellar halo metallicity $\sim$1~dex
lower than spirals of comparable luminosity (such as M31, as alluded to
earlier). 

However, it must be noted that the metallicity of the Galactic 
halo in the Solar Neighbourhood (e.g., in Ryan \& Norris 1991) comes 
from spectroscopic metallicities in a kinematically--selected 
sample, whereas that of the M31 halo, as well as those of the 
stellar halos of nearby spiral galaxies \citep{MM}, 
have been derived primarily from photometric metallicities in 
geographically--selected samples.  Because it is now becoming 
possible to obtain spectroscopic metallicities for significant 
samples of kinematically--selected giants in the M31 halo (e.g.: \citealt{Ibata04}; \citealt{Guhathakurta}), 
it will be crucial to assess the consistency of the M31 stellar halo metallicity as
derived from kinematically-- and from geographically--selected samples, respectively.

A grasp of the true scatter around the general trend in the halo metallicity -- luminosity relation awaits a larger observational data set, however any theory which attempts to explain such a relation needs to simultaneously account for the {\it apparent} metallicity discrepancy between the Milky Way and the Andromeda stellar halos. The question arises as to what is driving the scatter of halo metallicity for galaxies of comparable luminosity?

In what follows, we investigate whether differences in the stellar halo
assembly history can explain the diversity seen in halo metallicities.
Using chemo--dynamical numerical simulations,
we have constructed a sample 
of 89 model disc galaxies, spanning
$\sim$4~magnitudes in luminosity, and sampling a wide range of assembly histories
at a given luminosity (or mass).  We contrast the derived stellar halo
metallicity -- galactic luminosity relation with the recent empirical
determination of \cite{MM}. The numerical framework in which
the simulations have been conducted is described in Section~2, while
Sections~3 and 4 present the results and the related discussion, respectively.

\section{Simulations}

The simulations employed here are patterned after the semi--cosmological
adiabatic feedback model of \cite{Brook04a}, and were constructed using
the chemo--dynamical code {\tt GCD+} \citep{KG03}.  {\tt GCD+}
self--consistently treats
the effects of gravity, gas dynamics, radiative cooling, star
formation, and chemical enrichment (including Type~II and Ia Supernovae,
relaxing the instantaneous recycling approximation). 

This semi--cosmological version of \texttt{GCD+} is based on the
seminal work of \cite{KG91}.  The initial condition for a given
model is an
isolated sphere of dark matter and gas. This ``top--hat'' overdensity has an
amplitude $\delta_{i}$ at initial redshift $z_{i}$, which is approximately
related to the collapse redshift $z_{c}$ by $z_{c} = 0.36\delta_{i}(1 +
z_{i}) - 1$ (e.g., \citealt{Padmanabhan}). We set $z_{c} = 2.0$, which
determines $\delta_{i}$ at $z_{i} = 40$. Small--scale density fluctuations
based on a CDM power spectrum are superimposed on the sphere using
\texttt{COSMICS}\footnote{\tt http://arcturus.mit.edu/cosmics} 
(\citealt{Bertschinger}), and the amplitude of the fluctuations is
parameterised by $\sigma_{8}$. These fluctuations are the seeds for local
collapse and subsequent star formation. Solid--body rotation corresponding
to a spin parameter $\lambda$ 
is imparted to the initial sphere to incorporate the
effects of longer wavelength fluctuations. For the flat CDM model
described here, the relevant parameters include $\Omega_{0} = 1$, baryon
fraction $\Omega_{b} = 0.1$, $H_{0} = 50$~km~s$^{-1}$~Mpc$^{-1}$, spin
parameter $\lambda = 0.06$, and $\sigma_{8} = 0.5$.
We employed 14147 dark matter and 14147 gas/star
particles, for the 89 simulations presented here.

Using the same number of particles, collapse redshift $z_c$, and spin
parameter $\lambda$, a series of 89 simulations were completed spanning
a factor of 50 in mass in four separate mass ``bins'': M$_{\rm tot}$ = 1$\times$10$^{11}$~M$_\odot$;  5$\times$10$^{11}$~M$_\odot$; 
1$\times$10$^{12}$~M$_\odot$; 5$\times$10$^{12}$~M$_\odot$.
For each total
mass, we run models with different patterns of small--scale density
fluctuations which lead to different hierarchical assembly histories.  
This was controlled by setting different random seeds
for the Gaussian perturbation generator in {\tt COSMICS}.

\section{Results}

Our grid of 89 
simulations were employed to populate the redshift $z$ = 0
stellar halo metallicity -- luminosity plane 
(i.e. $\langle$[Fe/H]$\rangle_{\rm halo}$ -- M$_{\rm V}$)
shown in Fig.~1 (filled symbols\footnote{Note: MDF centroids were constrained
to sit on a 0.2~dex ``grid'' in $\langle$[Fe/H]$\rangle_{\rm halo}$,
which is the source of the apparently ``quantised'' centroid values in Fig.~1.}). We have been conservative in our
geometrical definition of ``stellar halo'', adopting a projected cut--off radius of 15~kpc
to delineate the halo from (possible) disc--bulge contaminants. Such a choice should also allow an easier comparison with the observations of halo fields in nearby spiral galaxies. 

The stellar
metallicity distribution function (MDF) was then generated using all stellar particles from the halo ``region'', convolved with a Gaussian of 
$\sigma$$_{\rm [Fe/H]}$ = 0.15~dex, representing the typical observational
uncertainties (e.g., \citealt{Bellazzini}).  The peaks of the derived MDFs are shown\footnote{Only those simulations with $>$100 halo stellar particles
are included in the analysis here.} in Fig.~1; 
68\% of the stellar particles within a given (simulated)
MDF typically span 1.4~dex in metallicity.  We
confirmed that the derived halo MDFs are
insensitive to projection effects, even in the most massive (largest) models, 
suggesting that our $R\sim$~15~kpc definition for the halo ``region'' is
sufficient for minimising disc and bulge contamination.  The optical
properties of our simulated stellar populations were derived using the 
population synthesis models of \cite{ML}, taking into account the age 
and the metallicity of each stellar particle.  We note that the V-band
luminosity shown along the abscissa of Fig.~1 is the {\it total} galactic
luminosity (i.e. halo + bulge + disc).

\begin{figure}
\begin{center}
\includegraphics[width=0.5\textwidth]{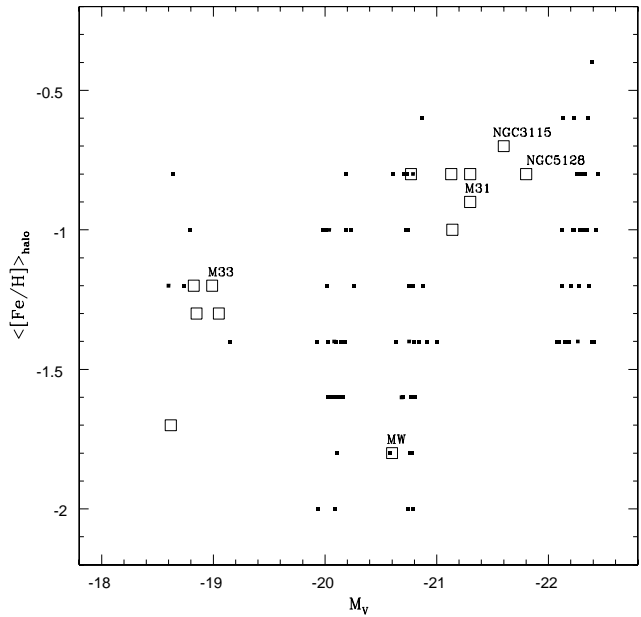}
\caption{The stellar halo metallicity -- total galactic V-band luminosity relation ($\langle{\rm [Fe/H]}\rangle_{\rm halo}$ -- M$_{\rm V}$).
Filled squares correspond to the peak of the MDF for each of our simulations. 
Open (unlabelled) squares represent the data of Mouhcine et~al. (2005); open (labelled) squares represent
additional data taken from the literature (see the text for details).  
68\% of the stars in the simulated (observed) MDFs are typically enclosed 
within $\pm$0.7~dex ($\pm$0.35~dex) of the peak of the respective MDF.}
\end{center}
\end{figure}

Fig.~1 also shows the corresponding observed halo
metallicity -- luminosity values for 13 nearby spirals, taken primarily from 
\cite{MM}, supplemented with data from the literature
(\citealt{BM}; \citealt{Elson}; Harris, Harris \& Poole 1999; 
Brooks,~Wilson~\&~Harris~2004)\footnote{Note that the identification of the stellar halo in M33 is still debated (e.g., Tiede, Sarajedini \& Barker 2004).}. For the observational
datasets, we calculated the MDF peak and 68\% ``dispersions'' exactly as we did
for the simulated datasets\footnote{Each MDF has been fitted by an 
univariate skew--normal distribution (e.g.: \citealt{Azzalini}; {\tt http://azzalini.stat.unipd.it/SN/Intro/intro.html}). For the observational datasets, we calculated the MDF as derived from the mean location of the Red Giant Branch stars in the Colour Magnitude Diagrams of the fields observed in \cite{MM}.}; 68\% of the stars in the observed MDFs are typically enclosed within
$\pm$0.35~dex of the MDF peak, a factor of $\sim$2 narrower than the
simulations\footnote{Our current model does not take into account any pre--enrichment scenario due to extremely metal--poor stars (Pop~III hereafter) whose detailed physics is still much debated (e.g.: Woosley, Heger \& Weaver 2002; \citealt{Larson}). An early and homogeneous pre--enrichment of the proto--galactic masses due to Pop~III could lead to narrower MDFs at lower redshift.}.

What is readily apparent from Fig.~1 is that significant variation in halo
metallicity ($\simgt$1~dex) exists at any given total galactic luminosity in our
simulations. For a set of models with the same initial total mass,
the only difference among these models can be traced
to the random pattern of initial small-scale density fluctuations;
this translates directly into differing hierarchical assembly histories.
Qualitiatively, it would appear that assembly history alone
may account for the diversity in halo metallicity at a given galactic luminosity,
and thus account for the apparent outliers in the observed trend.

\begin{figure}
\begin{center}
\includegraphics[width=0.5\textwidth]{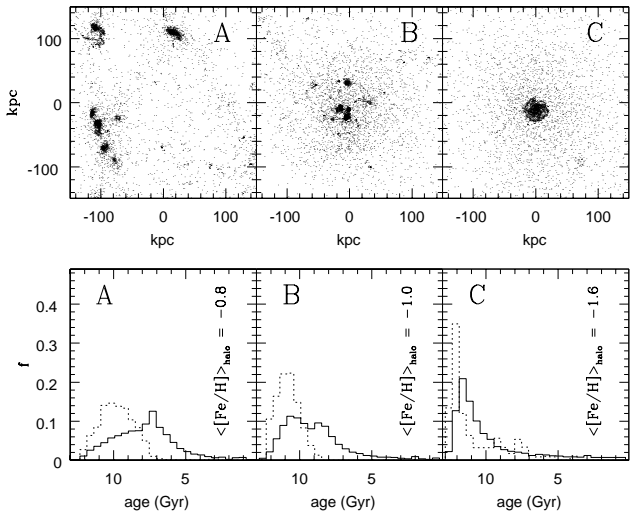}
\caption{\textit{Upper panels}: Projected distribution of the gas particles at
redshift $z = 1.5$ for the simulations with $M_{tot}$ = 10$^{12}$ M$_{\odot}$.
\textit{Lower panels}: The associated stellar age distributions at
$z = 0$ for the galaxies in the upper panels.
The solid (dotted) histogram corresponds to the
stellar age distribution function for the entire galaxy (stellar halo).
The corresponding halo metallicities are denoted in each panel.}
\end{center}
\end{figure}

Fig.~2 shows snapshots of the gas particles at redshift $z$ = 1.5 and the 
associated $z$ = 0 stellar Age Distribution Function (ADF) for
three of the models with M$_{\rm tot}$ = 10$^{12}$~M$_\odot$. 
These representative models demonstrate that the simulated galaxies with the
more metal--rich halos are assembled over a longer timescale,
and thus possess a broader ADF (both in the halo and in the associated galaxy).
Such a scenario is consistent with the evidence, presented by \cite{Brown},
of a substantial intermediate-age metal-rich population in a geographically-selected halo field
of M31\footnote{However, the eventual contamination (by the disc or by debris of accreted satellites)
in this outer field is debated (e.g.: \citealt{B03}; \citealt{Ferguson05}).}.
Conversely, the simulated galaxies with the more metal--poor halos
are assembled earlier through more of a monolithic process,
with a consequently narrower ADF.
Our simulations suggest that
the halo metallicity reflects directly the formation and assembly history
of the host galaxy\footnote{We note that the ``dispersion'' in the 
halo and total galactic ADFs appear correlated, such that galaxies with
narrower halo ADFs also appear to have narrower total galactic ADFs.
This could suggest that the assembly history of the halo ``knows'' of the
formation history of the other galactic structural components (e.g., bulge
and disc).  We will explore this suggestion in a future study.}.

\begin{figure}
\begin{center}
\includegraphics[width=0.5\textwidth]{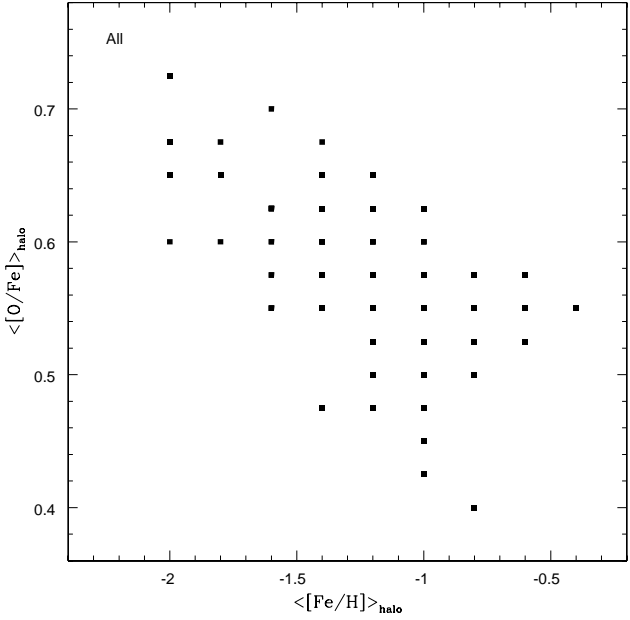}
\caption{The relation between the peak of the halo [O/Fe] distribution function and the peak of the halo MDF in our simulations.}
\end{center}
\end{figure}

Further, the most metal--poor stellar halos in our simulations (which, recall,
formed preferentially via more of a monolithic collapse) possess $\alpha$--elements to
iron ratios a factor of $\sim$2 higher than the most metal--rich halos (which
formed preferentially over more extended period of hierarchical clustering), 
as shown in Fig.~3. Such a trend is expected if the different amount of $\alpha$--elements and Iron released 
by Type~II and Ia Supernovae over different timescales is taken into account 
(e.g.: Timmes, Woosley \& Weaver 1995; \citealt{Woosley}).

\begin{figure}
\begin{center}
\includegraphics[width=0.5\textwidth]{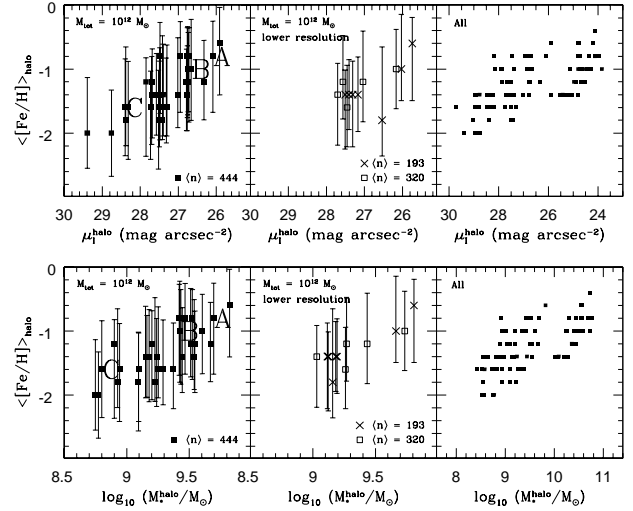}
\caption{\textit{Upper panels}: Stellar halo metallicity -- halo I-band surface brightness relation
($\langle$[Fe/H]$\rangle_{\rm halo}$ -- $\mu_{\rm I}^{\rm halo}$):
in the left panel, our fiducial models with
M$_{tot}$ = 10$^{12}$~M$_\odot$ and 14147$\times$2 particles are shown,
while the middle panel shows the corresponding lower--resolution
models with 5575$\times$2 (crosses) and 9171$\times$2 (open squares)
particles. The metallicity error bars indicate the 68\% Confidence Level.
The average number of halo particles is shown in the bottom
right corner of each panel.  Models with fewer than 100 stellar halo
particles (R$>$15~kpc) are not included here.  Models A~--~C of Fig.~2
are denoted in the left and middle panels. The right panel shows 
the $\langle$[Fe/H]$\rangle_{\rm halo}$ -- $\mu_{\rm I}^{\rm halo}$ relation
for all the simulations.
\textit{Lower panels}: Halo metallicity -- stellar mass relation
($\langle$[Fe/H]$\rangle_{\rm halo}$ -- M$_{*}^{\rm halo}$): in the left panel, our fiducial models with
M$_{tot}$ = 10$^{12}$~M$_\odot$, while the middle panel shows the corresponding lower--resolution
models. The right panel shows the $\langle$[Fe/H]$\rangle_{\rm halo}$ -- M$_{*}^{\rm halo}$ relation
for all the simulations.}
\end{center}
\end{figure}

The upper panels of Fig.~4 show the stellar halo metallicity -- I-band surface
brightness relation ($\langle{\rm [Fe/H]}\rangle_{\rm halo}$ -- $\mu_{\rm I}^{\rm halo}$) for models with 
M$_{\rm tot}$ = 10$^{12}$~M$_\odot$; the surface brightness was measured
at a projected distance of 20~kpc from the dynamical centre of each simulation. 
The halo metallicity -- stellar mass relation ($\langle$[Fe/H]$\rangle_{\rm halo}$ -- M$_{*}^{\rm halo}$)
is shown in the lower panels of Fig.~4. The three galaxies
presented in Fig.~2 are also labelled in Fig.~4. 
An immediate correlation is apparent with the more massive halos possessing higher
surface brightness and also higher metallicity.  This is consistent
with a picture in which galaxies that experienced more extended 
assembly histories have more massive stellar halos, with both higher halo metallicities and halo
surface brightnesses - i.e. higher stellar halo densities. Similar trends are observed in our other total mass models, 
as shown in the right panels of Fig.~4.

The issue of model convergence (resolution) is always a concern when
interpreting cosmological simulations (particularly when including
baryons).  To test this, we conducted a series of simulations for
M$_{\rm tot}$=10$^{12}$~M$_\odot$, with 5575$\times$2 and 9171$\times$2
particles, to supplement the default grid (which used 14147$\times$2
particles); the middle panels of Fig.~4 show where these lower--resolution
simulations sit in the halo metallicity -- surface brightness plane and in the halo metallicity -- stellar mass plane, respectively.
The consistency between higher (left panels) and lower (middle panels)
resolution models is reassuring and leads us to conclude that our 
simulations are not affected significantly by numerical 
resolution.\footnote{Because we limit our analyses to models with
$>$100 halo stellar particles, the $\mu_{\rm I}^{\rm 
halo}$$>$28~mag~arcsec$^{-2}$ region of the halo metallicity -- surface
brightness plane and the M$_{*}^{\rm halo}$$<$10$^{9}$M$_{\odot}$ region of the halo metallicity -- stellar mass plane 
are underpopulated by the lower--resolution models (middle
panels of Fig.~4).}

\section{Summary and Conclusions}
We have presented here an analysis of the characteristics of spiral galaxy 
stellar halos formed within a large grid of numerical simulations, with 
particular emphasis placed 
upon the relationship between stellar halo metallicity
and the associated galactic luminosity.  We have demonstrated that
at any given total luminosity (or, conversely, total dynamical mass),
the metallicities of these simulated
stellar halos span a range in excess of $\sim$~1~dex.  We suggest that the 
underlying driver of this metallicity spread can be traced to the 
diversity of galactic mass assembly histories inherent within the 
hierarchical clustering paradigm.  Galaxies with a more protracted 
assembly history possess more metal--rich and younger stellar halos, with
an associated greater dispersion in age, than galaxies which experience
more of a monolithic collapse.

For a given total luminosity (or dynamical mass), those galaxies with more
extended assembly histories also possess more massive stellar halos, which 
in turn leads to a direct correlation between a stellar halo's
metallicity and its surface brightness (as anticipated by earlier 
semi--analytical models - e.g., \citealt{Renda}).  By extension, such a 
correlation may prove to be a useful diagnostic tool for disentangling the
formation history of disc galaxies.  
 
Recently, \cite{MM} have presented an observed correlation between
stellar halo metallicity and total galactic luminosity, as shown in Fig.~1. 
The observed dispersion in the mean halo metallicity at a given galactic
luminosity is {\it smaller} than what we find in our simulations. {\it However}
the latter can account for the outliers in the observed trend. Since our motivation has been to study
which is the effect of the pattern of the initial density fluctuations {\it alone}
on the stellar halo features at redshift zero in simulated spiral galaxies,
it is worth to note that galaxy formation, as it is observed,
is an ongoing process which is the result of the interplay among different parameters,
of which the pattern of initial density fluctuations (thus the merging history) is one.
We have shown that the merging history {\it alone} may be held responsible of the dispersion
in halo metallicity at a comparable total galactic luminosity,
as {\it apparently} observed for example in our Milky Way and in Andromeda (see Section~1).

This begs the question...\textit{Which is normal}? 
Our study suggests that if the stellar halo was assembled (primarily)
through more of a monolithic collapse, such a low metallicity is indeed what should
be expected; conversely, the fact that the M31 stellar halo is significantly metal--rich is suggestive of a more
protracted assembly history.
An observational consequence of these differing formation histories is
the prediction that the M31 stellar halo should possess a high
surface brightness;
observations tentatively support this prediction (Reitzel, Guhathakurta \& Gould 1998; \citealt{Guhathakurta}; \citealt{Irwin}).

Further observations are needed to tighten our grasp of the strength and scatter of the stellar halo metallicity -- luminosity relation, which we have shown to be a useful diagnostic tool for disentangling the formation history of disc galaxies.

\section{Acknowledgements}

We thank the anonymous reviewer for careful comments which improved the manuscript. 
We acknowledge the financial support of the Australian Research Council
through its Discovery Project and Linkage International schemes, and the
Australian and Victorian Partnerships for Advanced Computing.
AR acknowledges the hospitality of the Observatoire Astronomique de Strasbourg 
and of the Tuorlan Observatorio, 
and the Swinburne Supercomputer Facility support team.
DK acknowledges the financial support of the Japan Society for the Promotion of Science, through a Postdoctoral Fellowship for research abroad.

\bsp
\label{lastpage}

\end{document}